\documentclass[12pt]{article}
\usepackage{epsfig}
\usepackage{amsmath}
\usepackage{amsfonts}
\usepackage{xspace}

\newcommand{\bea}{\begin{eqnarray}}
\newcommand{\eea}{\end{eqnarray}}
\newcommand{\bi}{\begin{itemize}}
\newcommand{\ei}{\end{itemize}}

\newcommand{\eq}{=}

\newcommand{\ud}{\text{d}}

\newcommand{\abs}[1]{\lvert#1\rvert}
\newcommand{\df}{\mathrm{d}}
\newcommand{\g}{\gamma}

\newcommand{\hm}{\widehat{m}}

\newcommand{\hF}{\widehat{F}}
\newcommand{\hW}{\widehat{W}}
\newcommand{\GeV}{\,\mathrm{GeV}}
\newcommand{\nn}{\nonumber}

\newcommand{\incl}{\mathrm{incl}}
\newcommand{\babar}{\mbox{\ensuremath{{\displaystyle B}\!{\scriptstyle A}
{\displaystyle B}\!{\scriptstyle AR}}}\xspace}
\newcommand{\belle}{\mbox{\emph{Belle}}\xspace}

\def\beq{\begin{equation}}
\def\eeq#1{\label{#1}\end{equation}}
\def\eeqn{\end{equation}}

\def\beqa{\begin{eqnarray}}
\def\eeqa#1{\label{#1}\end{eqnarray}}
\def\eeqan{\end{eqnarray}}

\let\bar=\overbar

\def\Dslash{\not{\hbox{\kern-4pt $D$}}}
\def\dslash{\not{\hbox{\kern-2pt $\del$}}}

\def\msb{{\bar{\ssstyle M \kern -1pt S}}}

\def\Title#1{\begin{center} {\Large {\bf #1} } \end{center}}

\begin{document}

\Title{A model independent determination of the $B \to X_s \gamma$ decay rate }

Proceedings of CKM 2012, the 7th International Workshop on the CKM Unitarity Triangle, University of Cincinnati, USA, 28 September - 2 October 2012\\ 

\bigskip\bigskip


\begin{raggedright}  

{\it Florian U. Bernlochner \\
              University of Victoria, Victoria, British Columbia, Canada V8W 3P\\
              E-Mail: florian@slac.stanford.edu}
\bigskip

{\it Heiko Lacker\\
              Humboldt University of Berlin, 12489 Berlin, Germany\\
              E-Mail: lacker@physik.hu-berlin.de}
              
\bigskip
{\it Zoltan Ligeti \\
              Ernest Orlando Lawrence Berkeley National Laboratory,
              University of California,\\ Berkeley, CA 94720, USA\\
              E-Mail: ligeti@lbl.gov}
              
\bigskip
{\it Iain W. Stewart \\
              Center for Theoretical Physics, Massachusetts Institute of Technology,\\Cambridge, MA 02139, USA\\
              E-Mail: iains@mit.edu}
              
\bigskip
{\it Frank J. Tackmann \\
              Deutsches Elektronen-Synchrotron (DESY), D-22607 Hamburg, Germany\\
              E-Mail: frank.tackmann@desy.de}
\bigskip

{\it Kerstin Tackmann \\
              Deutsches Elektronen-Synchrotron (DESY), D-22607 Hamburg, Germany\\
              E-Mail: kerstin.tackmann@desy.de}

\bigskip

{\it MIT-CTP 4429, DESY 13-019}

\bigskip\bigskip
\end{raggedright}

\section{Introduction}

\begin{figure}[t!]
\vspace{-4.5ex}
\hfill\includegraphics[width=0.4\textwidth]{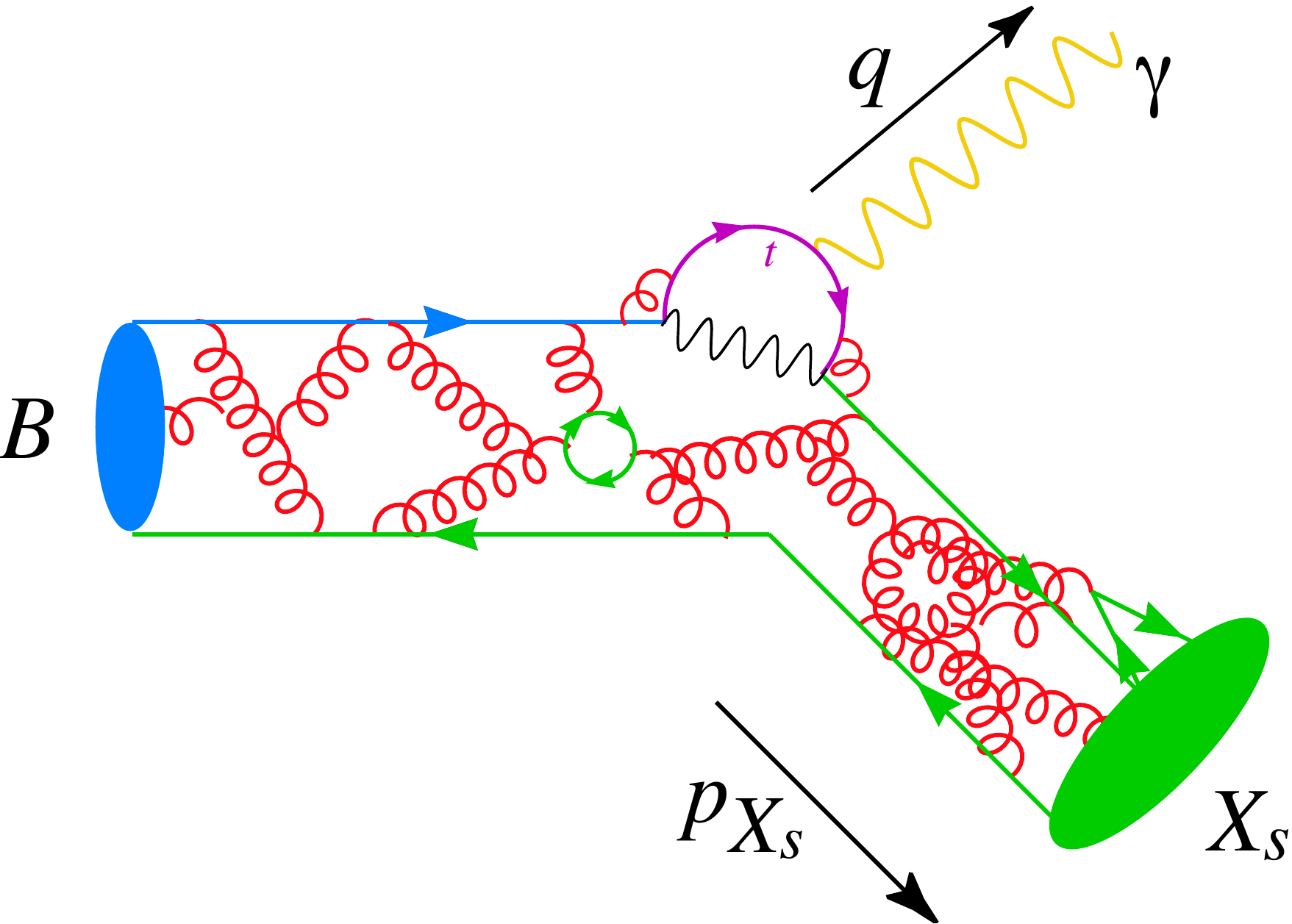}
\hfill\includegraphics[width=0.4\textwidth]{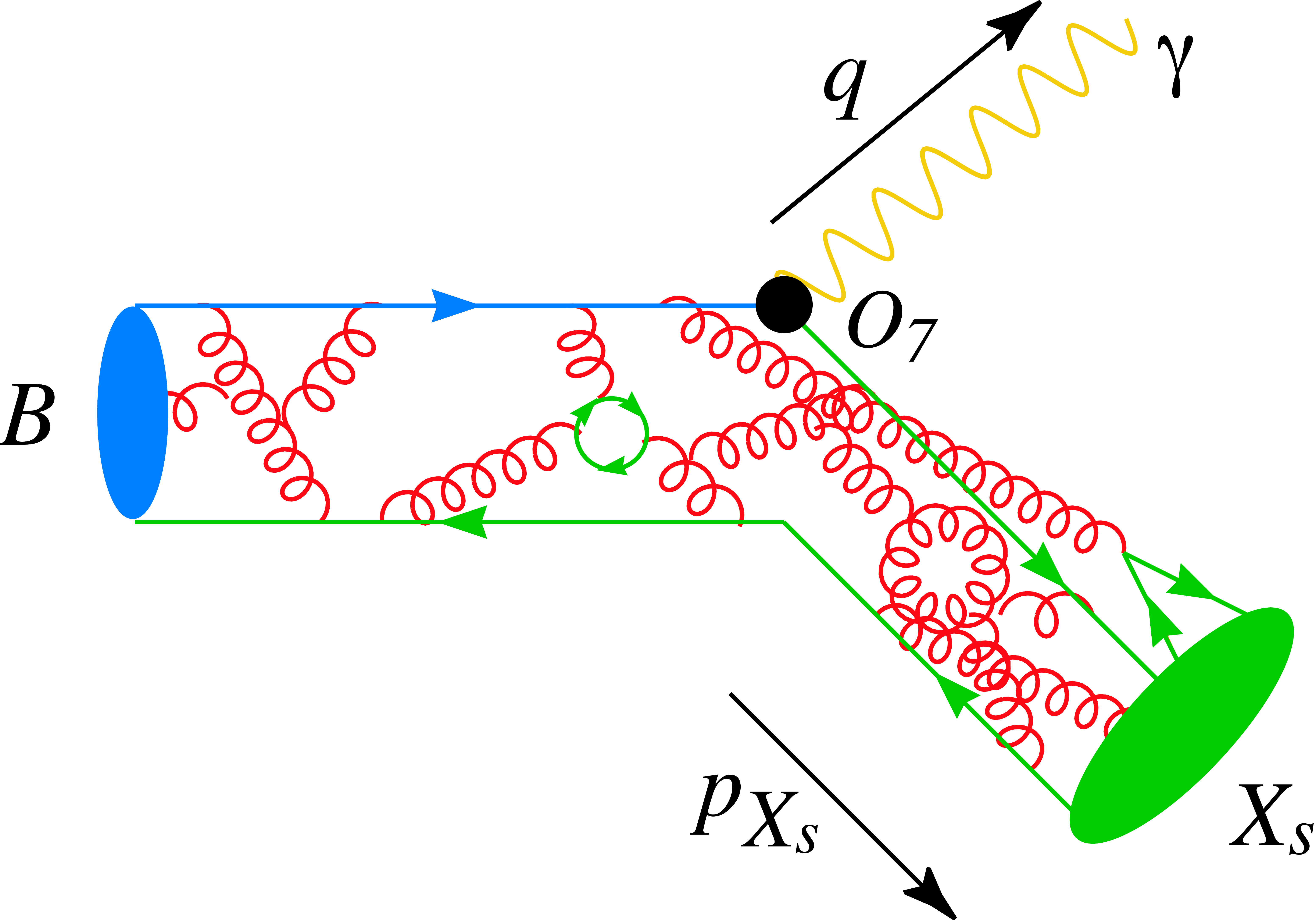}

\hspace*{\fill}
\caption{On the left-hand side the SM penguin $B \to X_s\, \g$ decay is shown. In new-physics models, such as e.g. the minimal supersymmetric Standard Model, the Wilson coefficient associated with the effective operator $\mathcal{O}_7$ shown at the right-hand side is modified.}
\label{chhiggs}
\vspace{-1ex}
\end{figure}

Rare penguin decays play an important role in the search for new physics in the flavor sector: the presence of new heavy particles which couple to heavy quarks would manifest itself in modifications of the total decay rate. In this presentation, a model-independent analysis is carried out to test the compatibility of four $B \to X_s \, \g$ measurements with the Standard Model expectation (SM). In Fig. \ref{chhiggs} the SM decay and a possible extension via a charged Higgs boson are shown. In the past, such an analysis used the extrapolation of the measured decay rates to the low $E_\g$ region to compare the experimental measured partial branching fraction with the next-to-next-to-leading order SM prediction, $\mathcal{B}(E_\g > 1.6 \, \text{GeV}) = \left( 3.55 \pm 0.24 \pm 0.09 \right) \times 10^{-4}$, from Ref.~\cite{Misiak:2006ab,Misiak:2006zs}, as adopted e.g. by the analysis of Ref.~\cite{Barberio:2006bi} and more recent updates. The reason for this extrapolation lies in the theoretically poorly known nonperturbative corrections from the $b$ quark distribution function in the $B$ meson, called the shape function, which affects the partial decay rate, $\mathcal{B}(E_\g > E_\g^{\text{cut}})$, for high values of $E_\g^{\text{cut}}$. The drawback of the extrapolation based analyses lies in the introduction of undesired model dependencies, which are hard to quantify, and also in making poor use of the experimentally most precise regions at high $E_\g$ by including the partial branching fraction down to the region of phase space with low $E_\g $, which is dominated by large background contributions from other $B$ meson decays. The SIMBA collaboration uses an alternative approach, outlined in Ref.~\cite{Ligeti:2008ac}, by determining the shape function directly from data from a global analysis of the $B \to X_s \, \g$ spectra with negligible model dependence compared to the present experimental and theoretical uncertainties. The main changes with respect to our earlier work Ref.~\cite{Bernlochner:2011di,Bernlochner:2010zz} lies in the evaluation of the uncertainties of missing higher-order perturbative corrections, and the inclusion of the measured $E_\gamma$ spectrum of Ref.~\cite{:2012iwb}.

\section{Treatment of the shape function and formulae for $B \to X_s \g$ decay rate}

\subsection{Shape function}

The shape function, which enters in the differential $B \to X_s \, \g$ decay rate, can be factorized into perturbative and non-perturbative contributions, cf. \cite{Ligeti:2008ac}, via
\begin{align}
  S(\omega , \mu) & \eq \int \ud k \, \, \widehat C_0(\omega-k, \mu) \,\, \widehat F(k) \, , \label{smastform}
\end{align}
where $\widehat C_0(\omega-k, \mu)$ is the $\overline{\text{MS}}$-renormalized $b$-quark matrix element of the shape function operator calculated in perturbation theory, and $\widehat F(k)$ are the non-perturbative contributions to $ S(\omega , \mu) $. Constructing the shape function as done in Eq.~\ref{smastform} offers several advantages over alternative approaches: the shape function has the correct perturbative tail at large $\omega$, and the correct RGE behavior; for small $\omega$, the shape function is dominated by the non-perturbative parameter $\widehat F(k)$. This means that the shape of the $B \to X_s \g$ spectrum at large $E_\g$ is determined by the non-perturbative parameter $\widehat F(k)$. 

The approach outlined in Ref.~\cite{Ligeti:2008ac} proposed the determination of $\widehat F(k)$ directly from experimental spectra by employing an expansion in a set of complete orthonormal basis functions, $f_n$, as
\begin{align} \label{eq:sfmastform}
 \widehat F(k) & \eq \tfrac{1}{\lambda} \left[   \sum_{n=0}^\infty \, c_n \, f_n \left( \tfrac{k}{\lambda} \right) \right]^2 \qquad \text{with} \qquad \int \ud k \, \widehat F(k) \quad \eq \quad \sum_{n=0}^\infty \, c_n^2 \quad \eq 1 \, ,
\end{align}
where $\lambda \sim \Lambda_{\text{QCD}}$ is a dimensional parameter of the basis. Since the orthonormal basis in Eq.~\ref{eq:sfmastform} is complete, this description offers a model-independent description of the shape function and the $B \to X_s \g$ decay rate, and the shape of the spectrum is parametrized by the expansion coefficients $c_n$. These coefficients can be determined directly by fitting the available data, taking into account the full experimental uncertainties and correlations. In practice, however, the expansion in Eq.~\ref{eq:sfmastform} needs to be truncated, since the available experimental information only allows for a finite number of $N$ coefficients to be constrained. This truncation introduces a residual model-dependence, which depends on the chosen functional basis and scales as $1 - \sum_{n=0}^N c_n^2$. The optimal values for $\lambda$ and $N$ need to be determined from data: $\lambda$ is chosen such that the fitted series converges quickly, and the number of used basis functions $N$ should be chosen large enough such that the truncation uncertainty is small compared to the experimental uncertainties. 

\subsection{Master formulae for $B \to X_s \g$ decay rate}

The $B \to X_s \g$ photon energy spectrum is given by 
\begin{align} \label{eq:master}
\frac{ \df \Gamma}{\df E_\g}
&= \frac{G_F^2 \alpha_\mathrm{em}}{2\pi^4}\, E_\g^3\, \hm_b^2\,\lvert V_{tb} V_{ts}^* \rvert^2
\nn\\ &\quad\times
\biggl\{
\abs{C_7^\incl}^2 \biggl[\int\! \df k\, \hW_{77}(k) \hF(m_B - 2E_\g - k)
+ \sum_m  \hW_{77,m}\, \hF_m(m_B - 2E_\g) \biggr]
\nn\\ &\qquad
+  \int\! \df k \sum_{i,j\neq7}  \Bigl[2\mathrm{Re}(C_7^\incl) C_i\, \hW_{7i}(k)
+ C_i C_j\, \hW_{ij}(k) \Bigr] \hF(m_B - 2E_\g - k)
\biggr\}
\,,\end{align}
where $\widehat W_{77}(k)$ contains the perturbative corrections to the $b \to s \g$ decay via the electromagnetic dipole operator, $O_7$, resummed to next-to-next-to-leading-logarithmic order \cite{Becher:2006pu, Ligeti:2008ac}, and including the full NNLO corrections \cite{Melnikov:2005bx, Blokland:2005uk}\footnote{At lowest order in perturbation theory: $\widehat W_{77}(k) = \delta (k)$}; $\widehat F(k)$ is the leading shape function as introduced in the previous section, the $\widehat F_m(k)$ denote $1/\widehat m_b$ suppressed subleading shape functions, and $\widehat W_{7i}(k)$ and $\widehat W_{ij}(k)$ contain subleading perturbative corrections. The full set of expressions entering Eq.~\ref{eq:master} will be given in Ref.~\cite{Ligeti:2010}. In a fit to $B \to X_s \g$ spectra the subleading shape functions can be absorbed at lowest order in $\alpha_s$ into the leading shape function, reducing the number of coefficients which need to be determined from data. The coefficient $C_7^\incl$ multiplying the dominant contribution proportional to $\widehat W_{77}(k)$ in Eq.~\ref{eq:master} is defined as
\begin{align} \label{eq:C7incl}
C_7^\incl
&= C_7^\mathrm{eff}(\mu_0) \frac{\overline m_b(\mu_0)}{\hm_b} + \sum_{i=1}^6 r_i(\mu_0)\, C_i(\mu_0)
+ r_8(\mu_0)\, C_8^\mathrm{eff}(\mu_0) \frac{\overline m_b(\mu_0)}{\hm_b}
+ \dotsb
\,,\end{align}
where $C_i^\mathrm{eff}(\mu_0)$ are the standard scheme-independent effective Wilson coefficients and $\overline m_b(\mu_0)$ is the $\overline{\mathrm{MS}}$ $b$-quark mass. The coefficients $r_{1-6,8}(\mu_0)$ contain all virtual contributions from the operators $O_{1-6,8}$ that generate the same effective $b\to s\g$ vertex as $O_7$. The ellipses denote included terms proportional to $\ln(\mu_0/\hm_b)$ that are required to cancel the $\mu_0$ dependence on the right-hand side and vanish at $\mu_0 = \hm_b$, such that $C_7^\mathrm{incl}$ is by definition $\mu_0$-independent to the order one is working at.

Since the terms in the last line in Eq.~\ref{eq:master} are small, we consider $\lvert C_7^\mathrm{incl}\, V_{tb} V_{ts}^* \rvert$ as the parameter that determines the normalization of the $B\to X_s\g$ rate. This normalization is extracted simultaneously with $\hF(k)$ from a fit to the various measured $E_\g$ spectra. The important contributions from $O_{1-6,8}$ are the virtual corrections contained in $C_7^\incl$, which have a sizable effect on the normalization of the $B\to X_s\g$ rate. By including them in $C_7^\incl$, they explicitly do not affect the shape of the spectrum, and so do not enter in our fit. They instead enter in the SM prediction for $C_7^\incl$, which can be computed independently. Below, the fit result is compared to the NLO SM value, $C_7^\incl = 0.354^{+0.011}_{-0.012}$~\cite{Ligeti:2010}. For a more stringent test for new physics, evaluating $C_7^\incl$ in the SM at NNLO along the lines of Refs.~\cite{Misiak:2006ab, Misiak:2006zs} would prove very valuable.

\section{Fit to the available $B$-Factory data}

\subsection{Experimental data from \babar and \emph{Belle}}

\begin{figure}[t!]
\vspace{-4.5ex}
\hspace{-12ex}\hfill\includegraphics[width=0.43\textwidth]{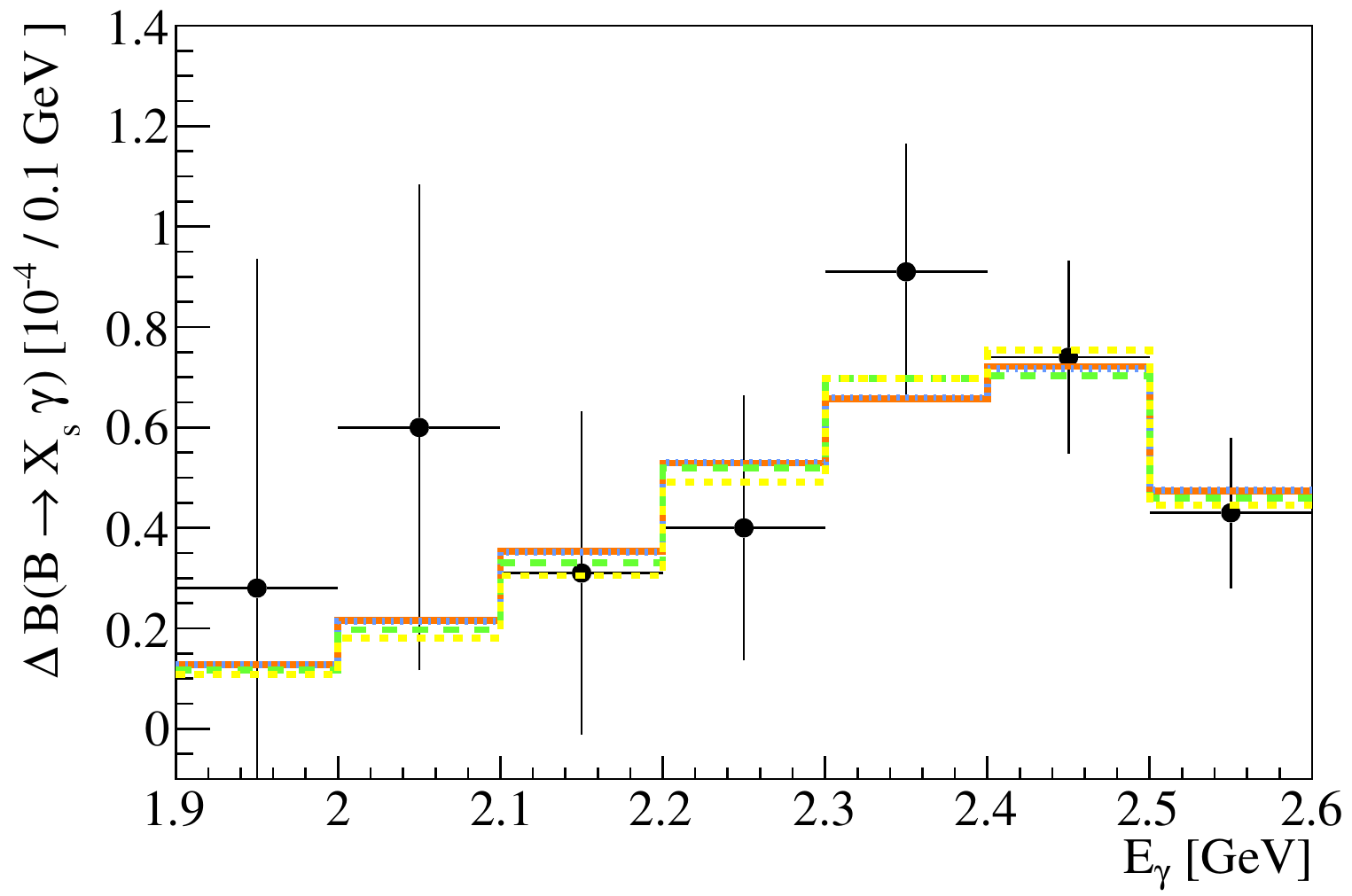}%
\hfill\includegraphics[width=0.43\textwidth]{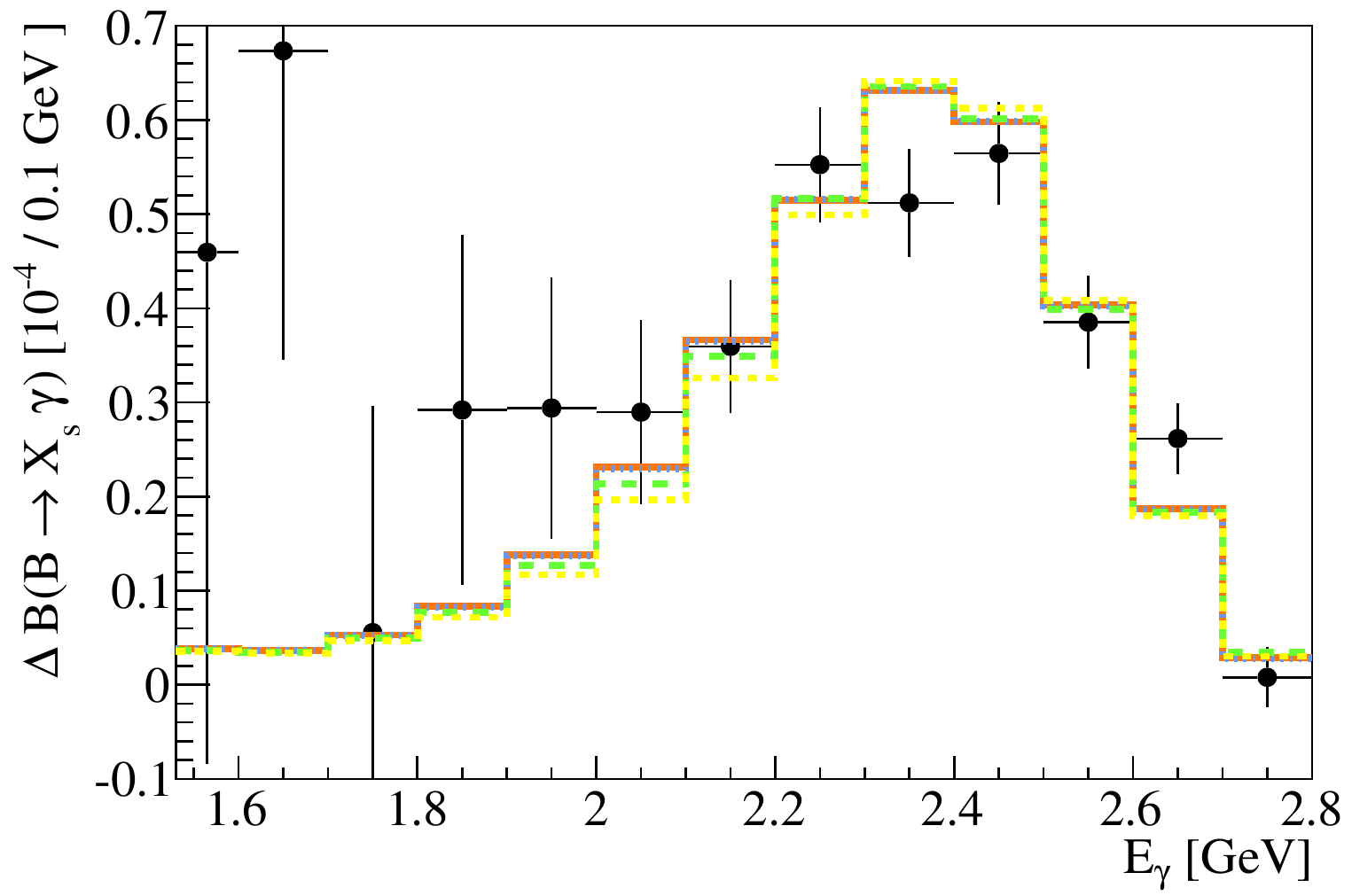} \\
\hfill\includegraphics[width=0.43\textwidth]{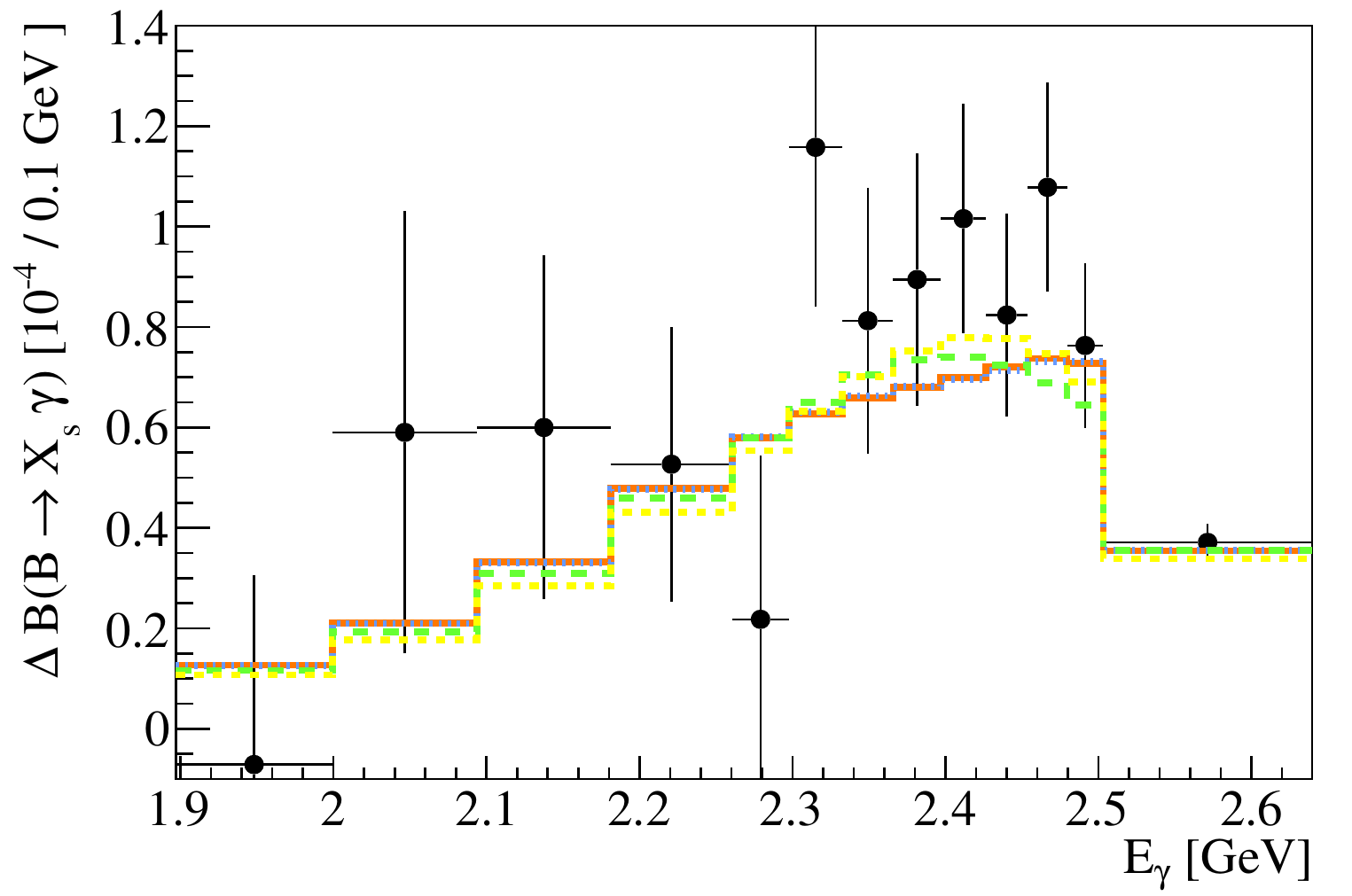}%
\hfill\includegraphics[width=0.43\textwidth]{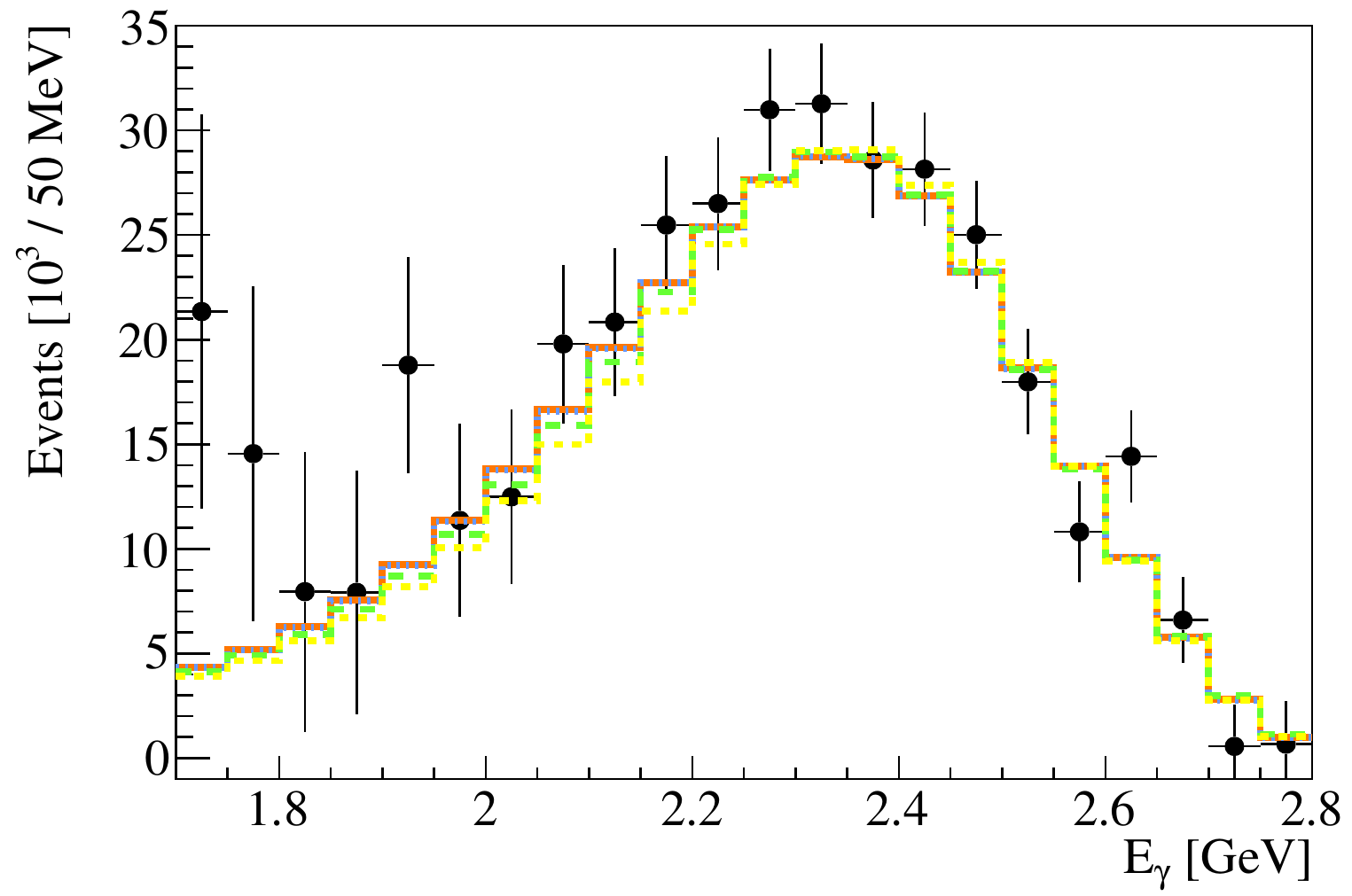} \\
\hspace*{\fill}
\vspace{-1ex}
\vspace{-4.0ex}

\caption{ Used \belle and \babar $B \to X_s \g$ measurements: Ref.~\cite{Aubert:2007my} (top left), Refs.~\cite{:2012iwb} (top right), Ref.~\cite{Aubert:2005cua} (bottom left), Ref.~\cite{:2009qg} (bottom right); The histograms show the result of the fit with a basis of $\lambda = 0.5$ GeV with two (yellow), three (green), four (blue), and four (orange) coefficients. The default fit result uses four coefficients. } \label{fitresult}
\vspace{-3ex}
\end{figure}

As experimental inputs the \belle measurement from Ref.~\cite{:2009qg}, and the three \babar measurements from Refs.~\cite{Aubert:2007my,Aubert:2005cua,:2012iwb} are used. The experimental statistical and systematic uncertainties and correlations are fully included in the fit procedure. The \babar spectra of Refs.~\cite{Aubert:2007my,Aubert:2005cua} are measured in the $B$ rest frame and are corrected for efficiencies. The experimental resolution in $E_\g$ for each spectrum is smaller than its respective bin size, so both spectra can be directly used in the fit. The \belle spectrum from Ref.~\cite{:2009qg} is measured in the $\Upsilon(4S)$ frame and affected by both efficiency and resolution. The \babar result of Ref.~\cite{:2012iwb} provides resolution unfolded spectra with smaller bin-by-bin correlations in the $B$ and the $\Upsilon(4S)$ frame. We analyze the unfolded result in the $\Upsilon(4S)$ frame. All four spectra are shown in Fig.~\ref{fitresult}.

\subsection{Fit setup}

To fit to the experimentally measured photon energy spectra, the expansion for $\hF(k)$ in Eq.~\ref{eq:sfmastform} is used for Eq.~\ref{eq:master} and integrated over the appropriate range of $E_\g$ for each experimental bin and each combination of basis functions $f_m(x) f_n(x)$. The theory prediction for the $i$th bin, $B^i$, is then given by
\begin{equation} \label{eq:fit}
B^i = \hm_b^2\, \abs{C_7^\incl V_{tb} V_{ts}^*}^2 \sum_{m,n=0}^N c_m c_n B_{mn}^i + \dotsb
\,,\end{equation}
where the ellipses denote the additional included terms arising from the last line in Eq.~\ref{eq:master}. The overall $\hm_b^2$ is expressed in terms of the moments of $\hF(k)$, so it is effectively a function of the $c_n$ coefficients. A $\chi^2$ minimization is performed to all available bins with $c_{0,1,...,N}$ and $\abs{C_7^\incl V_{tb} V_{ts}^*}$ as the fit parameters. The constraint $c_0^2 + \dotsb + c_N^2 = 1$ is enforced to ensure that $\hF(k)$ is properly normalized to unity. This fitting procedure was extensively tested using pseudo-experiments and provides unbiased central values with correct uncertainties.

\subsection{Fit results}

\begin{figure}[t!]
\vspace{-4.5ex}
\parbox{0.5\textwidth}{\includegraphics[width=0.5\textwidth]{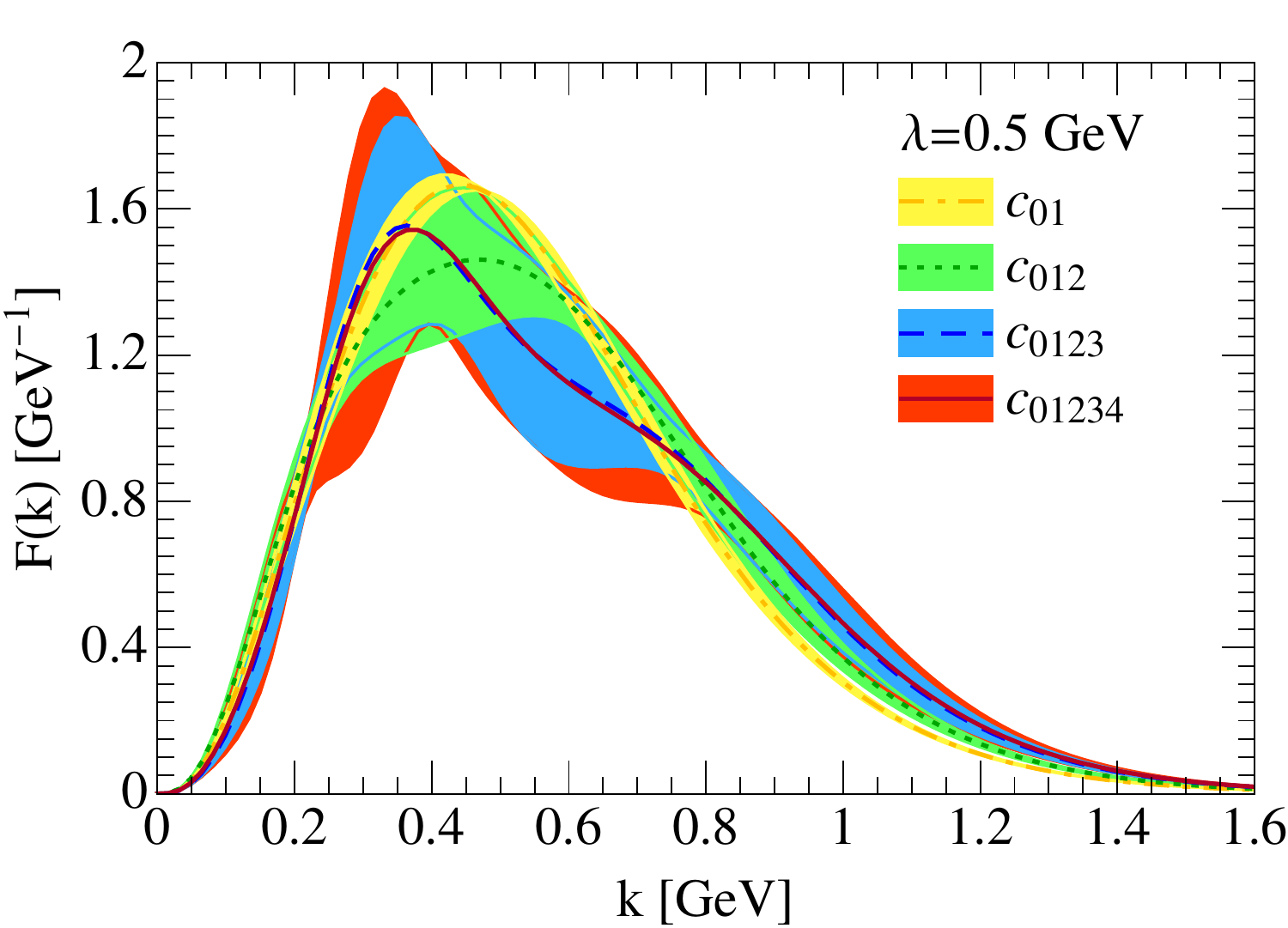}}%
\parbox{0.5\textwidth}{\includegraphics[width=0.5\textwidth]{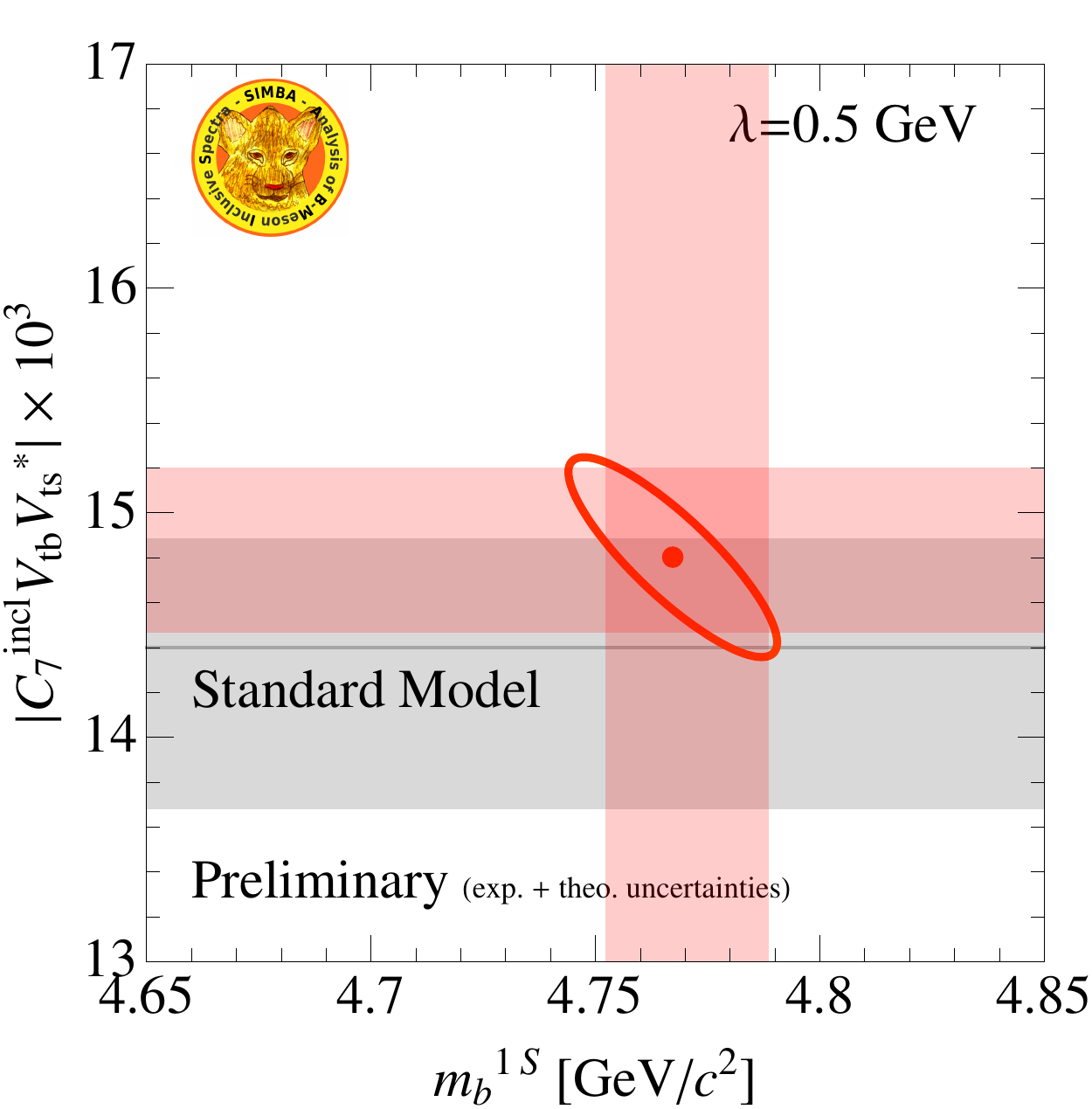}}
\vspace{-1ex}
\caption{The extracted $\widehat{F}(k)$ (with absorbed $1/m_b$ corrections) for two ($c_{0,1}$), three ($c_{0,1,2}$), four ($c_{0,1,2,3}$), and five ($c_{0,1,2,3,4}$) coefficients and basis parameter $\lambda = 0.5\GeV$ (Left). The colored envelopes are given by the uncertainties and correlations of the extracted coefficients $c_n$. Determined values of $\lvert C_7^\mathrm{incl} \, V_{tb} V_{ts}^{*}\rvert$ and $m_b^{1S}$ with $\Delta\chi^2 = 1$ contour (Right). The grey band shows the SM value of $\lvert C_7^\mathrm{incl} \, V_{tb} V_{ts}^{*}\rvert$ and the pink band show the SCET scale variation uncertainties.}   \label{SFC7mb}
\vspace{-1ex}
\end{figure}

For our default fit a value of $\lambda = 0.5$ GeV and four basis coefficients $c_{0,1,2,3}$ was chosen: These values were selected after checking carefully the convergence of various basis expansions, number of coefficients, and with the general focus of avoiding over-tuning. The corresponding fits with two $(c_{0,1})$, three $(c_{0,1,2})$, four $(c_{0,1,2,3})$, and five $(c_{0,1,2,3,4})$ expansion coefficients with  $\lambda = 0.5$ GeV are shown in Fig.~\ref{fitresult}. The fit converges after the inclusion of four coefficients and describes the measured spectra well. The $\chi^2 / \text{ndf}$ for the default fit with four coefficients is $41.65/48$ corresponding to a p-value of $0.87$. The fit results for the shape function for $2,3,4,$ and $5$ basis coefficients are shown in the left panel of Fig.~\ref{SFC7mb}. The corresponding results for $\abs{ C_7^\incl V_{tb} V_{ts}^{*}}$ and $m_b^{1S}$, where the latter is computed from the moments of the fitted $\hF(k)$, are shown in the right panel of Fig.~\ref{SFC7mb}. The shape function in Fig.~\ref{SFC7mb} verify the convergence of the basis expansion as the number of basis functions is increased. As one expects, the uncertainties returned by the fit increase with more coefficients due to the larger number of degrees of freedom. However, with too few coefficients one would have to add the truncation uncertainty. A reliable value for the final uncertainty is provided by the fitted uncertainty when the central values have converged and the respective last coefficients, here $c_3$ or $c_4$, are compatible with zero. At this point, the truncation uncertainty can be neglected compared to the fit uncertainties. Equivalently, the increase in the fit uncertainties from including the last coefficient that is compatible with zero effectively takes into account the truncation uncertainty. Using a fixed model function and fitting one or two model parameters would thus underestimate the true model uncertainties in the shape function model.

\subsection{Theoretical uncertainties from missing higher order perturbative corrections}


Including the theory uncertainties from missing higher order perturbation corrections for the default fit gives
\begin{align}
\abs{C_7^\incl V_{tb} V_{ts}^{*}} & = \bigl( 14.83 \pm 0.53_\mathrm{[exp]} \pm 0.37_\mathrm{[theo]} \bigr)\times 10^{-3} \, , \nonumber \\
m_b^{1S} & \eq 4.77 \pm 0.03_\mathrm{[exp]} \pm 0.02_\mathrm{[theo]}
\,,\end{align}
with comparable sizes for experimental and theoretical uncertainties. The result for $\abs{C_7^\incl}$ is compatible within one sigma with the NLO SM value, shown as a grey band in Fig.~\ref{SFC7mb}, for which a value of $\abs{V_{tb} V_{ts}^{*}} = 40.68^{+0.4}_{-0.5} \times 10^{-3}$ was used.

\section{Summary and outlook}

We presented preliminary results from a global fit to $B\to X_s\g$ data, which determines the total $B\to X_s\g$ rate, parametrized by $\abs{C_7^\incl V_{tb} V_{ts}^{*}}$, and the $B$-meson shape function within a model-independent framework. The value of $\abs{C_7^\incl V_{tb} V_{ts}^{*}}$ extracted from data agrees with the SM prediction within uncertainties. From the moments of the extracted shape function we determine $m_b^{1S}$. In the future, information on $m_b$ from other independent determinations can be included. The shape function extracted from $B\to X_s\g$ is an essential input to the determination of $\abs{V_{ub}}$ from inclusive $B\to X_u\ell\nu$ decays.

A combined fit to $B \to X_s \g$ and $B \to X_u\ell\nu$ measurements within our framework is in progress. It will allow for a simultaneous determination of $\abs{C_7^\incl V_{tb} V_{ts}^{*}}$ and $\abs{V_{ub}}$ along with the shape function with reliable uncertainties. In addition to a few branching fractions with fixed cuts, it is important to have measurements of the $B\to X_u\ell\nu$ differential spectra (including correlations), e.g.\ the lepton energy or hadronic invariant mass spectra. As for $B\to X_s\g$, fitting the differential spectra allows making maximal use of the $B\to X_u\ell\nu$ measurement, by using the experimental most precise regions to constrain the nonperturbative inputs and further reduce the associated uncertainties.

\section*{Acknowledgments}


We thank Kevin Flood, Alan Eisner, Kyle Knoepfel, Jessop Colin, John Walsh and their colleagues from \babar for providing us with essential information about Ref.~\cite{:2012iwb} and very useful discussions. We are grateful to Antonio Limosani from \belle for providing us with the detector response matrix of Ref.~\cite{:2009qg}. We thank Francesca Di Lodovico from \babar, who provided us with the experimental correlations of Ref.~\cite{Aubert:2005cua}. This work was supported in part by the Director, Office of Science, Offices of High Energy and Nuclear Physics of the U.S.\ Department of Energy under the Contracts
DE-AC02-05CH11231 (Z.L.) and DE-FG02-94ER40818 (I.S), and by the DFG Emmy-Noether Grant No. TA 867/1-1 (F.T.)..

\vspace{-1ex}
\bibliographystyle{jhep}

\bibliography{simba}

\providecommand{\href}[2]{#2}\begingroup\raggedright\begin{thebibliography}{10}

\bibitem{Misiak:2006ab}
M.~Misiak and M.~Steinhauser, {\it {NNLO QCD corrections to the $\bar B\to
  X_s\gamma$ matrix elements using interpolation in $m_c$}},  {\em Nucl. Phys.
  B} {\bf 764} (2007) 62--82, [\href{http://arXiv.org/abs/hep-ph/0609241}{{\tt
  hep-ph/0609241}}].

\bibitem{Misiak:2006zs}
M.~Misiak {\em et.~al.}, {\it {The first estimate of $B(\bar B \to X_s \gamma)$
  at $O(\alpha_s^2)$}},  {\em Phys. Rev. Lett.} {\bf 98} (2007) 022002,
  [\href{http://arXiv.org/abs/hep-ph/0609232}{{\tt hep-ph/0609232}}].

\bibitem{Barberio:2006bi}
{\bf Heavy Flavor Averaging Group (HFAG)} Collaboration, E.~Barberio {\em
  et.~al.}, {\it {Averages of $b-$hadron properties at the end of 2005}},
  \href{http://arXiv.org/abs/[hep-ex/0603003]}{{\tt [hep-ex/0603003]}}.

\bibitem{Ligeti:2008ac}
Z.~Ligeti, I.~W. Stewart, and F.~J. Tackmann, {\it {Treating the b quark
  distribution function with reliable uncertainties}},  {\em Phys. Rev. D} {\bf
  78} (2008) 114014, [\href{http://arXiv.org/abs/0807.1926}{{\tt
  arXiv:0807.1926}}].

\bibitem{Bernlochner:2011di}
F.~U. Bernlochner, H.~Lacker, Z.~Ligeti, I.~W. Stewart, F.~J. Tackmann, {\em
  et.~al.}, {\it {Status of SIMBA}},
  \href{http://arXiv.org/abs/1101.3310}{{\tt arXiv:1101.3310}}.

\bibitem{Bernlochner:2010zz}
F.~U. Bernlochner, H.~Lacker, Z.~Ligeti, I.~W. Stewart, F.~J. Tackmann, {\em
  et.~al.}, {\it {Towards a global fit to extract the $B\to X_s \gamma$ decay
  rate and Vub}},  {\em PoS} {\bf ICHEP2010} (2010) 229,
  [\href{http://arXiv.org/abs/1011.5838}{{\tt arXiv:1011.5838}}].

\bibitem{:2012iwb}
{\bf BABAR} Collaboration, J.~Lees, V.~Poireau, and V.~Tisserand, {\it
  {Measurement of B($B \to X_s \gamma$), the $B \to X_s \gamma$ photon energy
  spectrum, and the direct CP asymmetry in $B \to X_{s+d} \gamma$ decays}},
  \href{http://arXiv.org/abs/1207.5772}{{\tt arXiv:1207.5772}}.

\bibitem{Becher:2006pu}
T.~Becher and M.~Neubert, {\it {Analysis of $\mathcal{B}(B \to X_s \gamma)$ at
  NNLO with a cut on photon energy}},  {\em Phys. Rev. Lett.} {\bf 98} (2007)
  022003, [\href{http://arXiv.org/abs/hep-ph/0610067}{{\tt hep-ph/0610067}}].

\bibitem{Melnikov:2005bx}
K.~Melnikov and A.~Mitov, {\it {The photon energy spectrum in $B \to X_s
  \gamma$ in perturbative QCD through $\mathcal{O}(\alpha_s^2)$}},  {\em Phys.
  Lett.} {\bf B620} (2005) 69--79,
  [\href{http://arXiv.org/abs/hep-ph/0505097}{{\tt hep-ph/0505097}}].

\bibitem{Blokland:2005uk}
I.~R. Blokland, A.~Czarnecki, M.~Misiak, M.~Slusarczyk, and F.~Tkachov, {\it
  {The electromagnetic dipole operator effect on $\bar B \to X_s\gamma$ at
  $O(\alpha_s^2)$}},  {\em Phys. Rev. D} {\bf 72} (2005) 033014,
  [\href{http://arXiv.org/abs/hep-ph/0506055}{{\tt hep-ph/0506055}}].

\bibitem{Ligeti:2010}
Z.~Ligeti, I.~W. Stewart, and F.~J. Tackmann.
\newblock \emph{Manuscript in preparation}.

\bibitem{:2009qg}
{\bf Belle} Collaboration, A.~Limosani {\em et.~al.}, {\it {Measurement of
  Inclusive Radiative B-meson Decays with a Photon Energy Threshold of 1.7
  GeV}},  {\em Phys. Rev. Lett.} {\bf 103} (2009) 241801,
  [\href{http://arXiv.org/abs/0907.1384}{{\tt arXiv:0907.1384}}].

\bibitem{Aubert:2007my}
{\bf BABAR} Collaboration, B.~Aubert {\em et.~al.}, {\it {Measurement of the $B
  \to X_s \, \gamma$ Branching Fraction and Photon Energy Spectrum using the
  Recoil Method}},  {\em Phys. Rev. D} {\bf 77} (2008) 051103,
  [\href{http://arXiv.org/abs/0711.4889}{{\tt arXiv:0711.4889}}].

\bibitem{Aubert:2005cua}
{\bf BABAR} Collaboration, B.~Aubert {\em et.~al.}, {\it {Measurements of the
  $B \to X_s \gamma$ branching fraction and photon spectrum from a sum of
  exclusive final states}},  {\em Phys. Rev. D} {\bf 72} (2005) 052004,
  [\href{http://arXiv.org/abs/hep-ex/0508004}{{\tt hep-ex/0508004}}].

\end{thebibliography}\endgroup
\vspace{-1ex}

\end{document}